\documentclass[prb,preprint]{revtex4-1} 

\usepackage{amsmath}  
\usepackage{amsfonts} 
\usepackage{graphicx} 
\usepackage{hyperref}
\usepackage{varioref}
\usepackage{siunitx}
\usepackage{array} 
\usepackage{booktabs} 
\usepackage{tabularx} 
\usepackage{longtable} 
\usepackage{multirow} 
\usepackage{threeparttable} 
\usepackage{verbatim, listings} 
\usepackage[numbered,framed]{matlab-prettifier}
\usepackage{textcmds}

\newcommand{\fig}[1]{\textsc{FIG.}~\ref{#1}}
\newcommand{\eq}[1]{Equation~\ref{#1}}

\newcommand{\app}[1]{Appendix~\ref{#1}}
\newcommand{\sect}[1]{Section~\ref{#1}}

\newcommand{\MHz}[0]{\,\si{\mega \hertz}}
\newcommand{\Ohm}[0]{\,\si{\ohm}}
\newcommand{\mOhm}[0]{\,\si{\milli \ohm}}
\newcommand{\MOhm}[0]{\,\si{\mega \ohm}}

\newcommand{\s}[0]{\,\si{\second}}
\newcommand{\ns}[0]{\,\si{\nano \second}}
\newcommand{\us}[0]{\,\si{\micro \second}}

\newcommand{\mm}[0]{\,\si{\milli \metre}}
\newcommand{\cm}[0]{\,\si{\centi \metre}}
\newcommand{\m}[0]{\,\si{\metre}}
\newcommand{\km}[0]{\,\si{\kilo \metre}}

\newcommand{\V}[0]{\,\si{\volt}}
\newcommand{\nF}[0]{\,\si{\nano \farad}}
\newcommand{\uF}[0]{\,\si{\micro \farad}}
\newcommand{\um}[0]{\,\si{\micro \metre}}
\newcommand{\nC}[0]{\,\si{\nano \coulomb}}

\begin{document}


\title{The Cable to the Moon: Veritasium's Light Bulb Experiment in Low-Cost Miniature Form}

\author{Michael Lenz}
\email{michael.lenz@sankt-afra.de} 
\affiliation{Sächsisches Landesgymnasium St. Afra zu Meißen -- Hochbegabtenförderung, Fachschaft Physik, Freiheit 13, 01662 Meißen (Germany)}



\date{\today}

\begin{abstract}

In a popular YouTube video by the channel \qq{Veritasium}, the following question is posed: 

\textit{Imagine you have a giant circuit consisting of a battery, a switch, a light bulb, and two wires which are each $300{,}000 \km$ long. That is the distance that light travels in one second. So, they [the wires] would reach out halfway to the Moon and then come back to be connected to the light bulb, which is one meter away. Now the question is: \qq{After I close this switch, how long would it take for the light bulb to light up: Is it half a second, one second, two seconds, $1\m/c$ or none of the above?}}


As part of the \qq{Physics Specialist Camp 2022} in Seifhennersdorf -- a final event of the Saxon Physics Olympiad -- students from grades 9 and 10 built a miniature model consisting of a printed circuit board and two 10-meter cables to experimentally study the question. 

This article describes the experimental setup, presents some exemplary results, and explains them in detail. 
To enable reconstruction for educational purposes, the final PCB layout, as well as the supply sources and design considerations, are provided.
\end{abstract}

\maketitle 

\section{Introduction} 
\label{sec:Introduction}
\begin{figure}[htbp]
\centering
\includegraphics[width=10 cm]{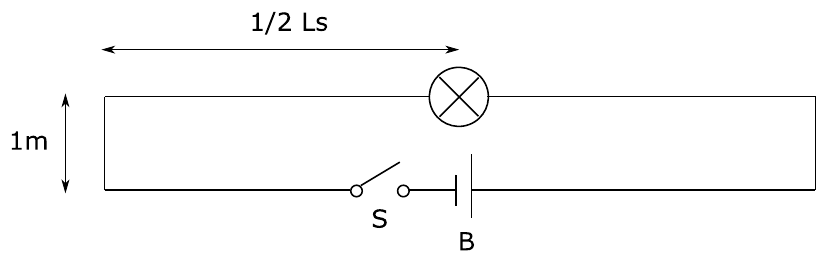}
\caption{Measurement setup in the Veritasium video, with a light bulb ($\otimes$), a switch (S), and a \hbox{battery (B)}, connected by two wires, each one light-second in length}
\label{fig:Schaltkreis}
\end{figure}

In his widely noticed video \citep{Veritasium}, the channel \q{Veritasium -- an element of truth} raises the question of how an electric light bulb with long wires (see \fig{fig:Schaltkreis}) is supplied with energy. The question is:  
\begin{itemize}
\item Do the electrons need to be set in motion along the entire length of the conductor, leading to a long delay between flipping the switch and the light bulb lighting up, or
\item does the light start shining much earlier, with the energy taking the direct path between battery and light bulb?
\end{itemize}

To study this question, a circuit board with the following elements was constructed as a miniature model:
\begin{itemize}
\item a DC voltage source of $2.5\V$ with low source impedance as a replacement for the battery, 
\item a switch based on a field-effect transistor and a gate driver for fast switching of the field-effect transistor (switching time approximately $10\ns$),
\item connections for two $10\m$ long coaxial cables to replace the one half light-second long cables to the left and right,
\item a load resistor to replace the light bulb,
\item three amplifier circuits to tap the voltages of interest at the switch and on both sides of the resistor, thereby decoupling the measurement cables from the circuit of interest and avoiding strong interferences.
\end{itemize}

\sect{sec:Circuit} presents the circuit used for the measurements. \sect{sec:Measurements} shows some exemplary measurement results, explains them based on the linear transmission line theory, and describes how to simulate the circuit in LTSpice. \sect{sec:Conclusion} finally summarizes the experiment and its outcome. If needed for your own teaching, \app{sec:Sources} provides the PCB description and the sources for the components making it possible to replicate and adapt the circuit. For a deeper understanding of the circuit, the design considerations are explained in \app{sec:Design-considerations}. Finally, \app{app:Voltage} provides a MATLAB script used to calculate the voltage across the load resistor.

\section{Measurement Circuit}
\label{sec:Circuit}

\begin{figure}[p]
\centering
\includegraphics[width=15 cm]{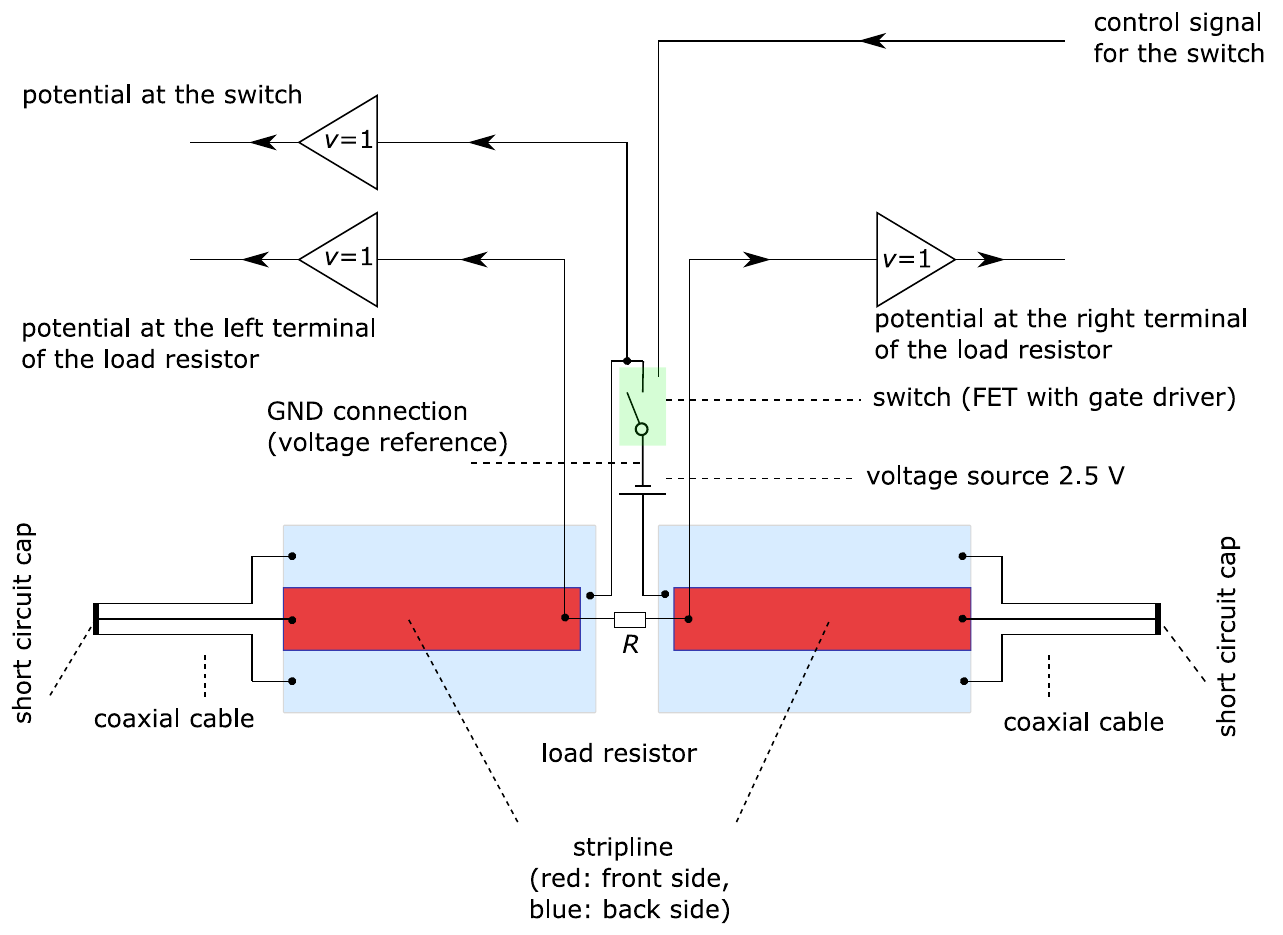}
\includegraphics[width=15 cm]{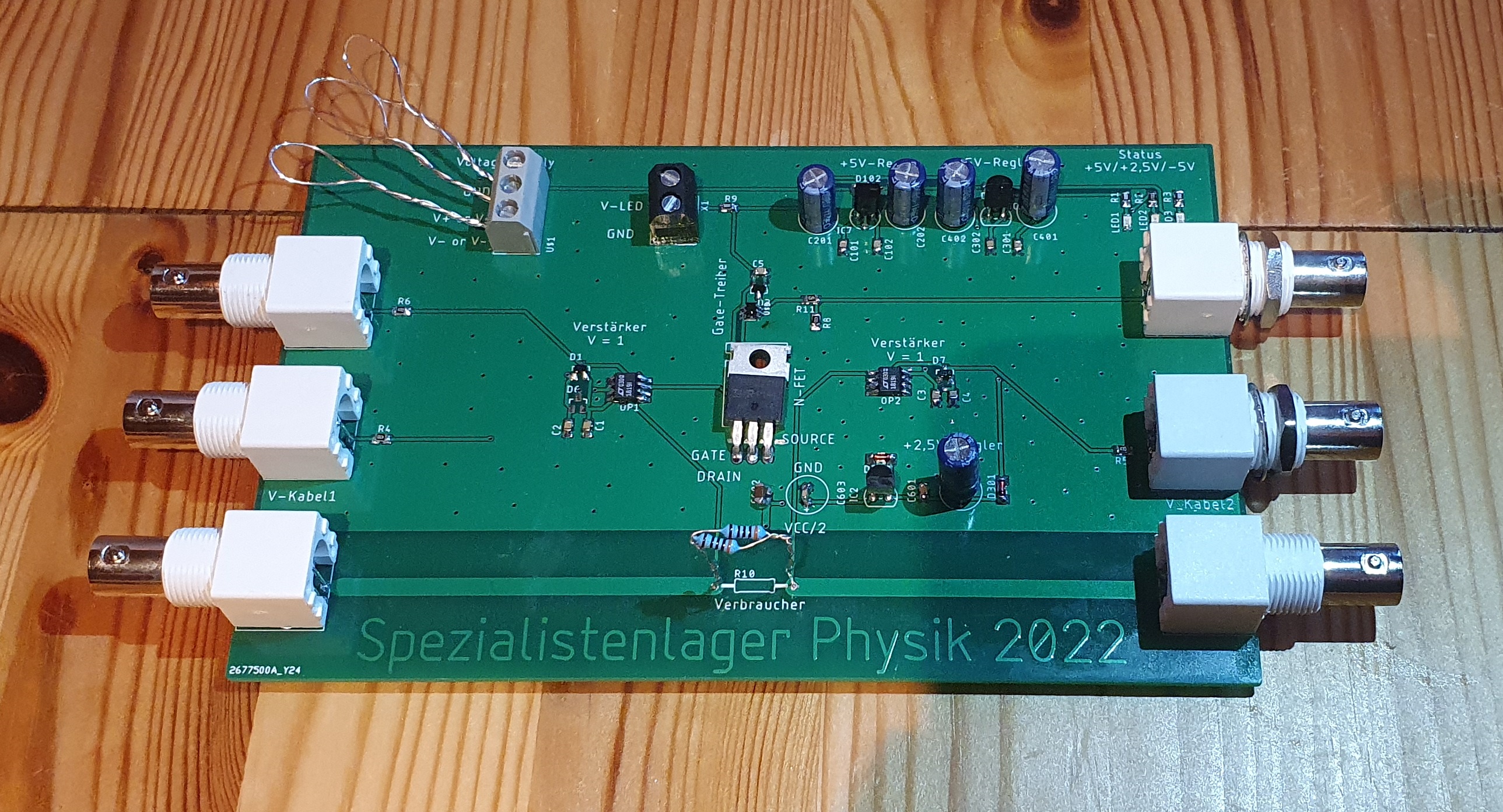}
\caption{Sketch of the measurement setup (top) and its technical realization as a circuit board built by students at the Physics Specialist Camp (bottom).}
\label{fig:Blockschaltbild}
\end{figure}

\fig{fig:Blockschaltbild} shows the main elements of the circuit used for the study: When used as a DC circuit, the current starts from the positive side of the constant voltage source ($2.5\V$) depicted in the center and flows from there through the bottom side (blue) of the right-hand stripline, the shielding of the right-hand coaxial cable, the short-circuit cap of the right-hand coaxial cable, the inner conductor of the right-hand coaxial cable, the top of the right-hand stripline (red), the load resistor, the top of the left-hand stripline (red), the inner conductor of the left-hand coaxial cable, the short-circuit cap of the left-hand coaxial cable, the outer conductor of the left-hand coaxial cable, the bottom of the left-hand stripline (blue) and back via the switch to the negative side of the voltage source.

A unity gain voltage amplifier is used to amplify each of the following measurement quantities:
\begin{itemize}
\item the potential at the switch (used to determine the exact switching time), 
\item the potential at the left terminal of the load resistor, and
\item the potential at the right terminal of the load resistor.
\end{itemize}

The idea is as follows: The difference between the potentials at the right and left terminals of the load resistor indicates the voltage across the load resistor. When the voltage is high, the electric power has reached the load resistor and the light bulb, respectively. Thus, a comparison between the voltage across the load resistor and the potential at the switch provides the delay time between flipping the switch and the light bulb lighting up.

\section{Measurements}
\label{sec:Measurements}

\subsection{Signal speed along the coaxial cable}
\label{sec:Lichtgeschwindigkeit}
To measure the signal speed along the coaxial cable, a signal generator was connected to one channel of an oscilloscope via a $30\cm$ short cable. Using a $10 \m$ coaxial cable of the same type, the signal was then transmitted to another channel of the same oscilloscope and terminated with a $50 \Ohm$ resistor to avoid reflections. The measurement was carried out with a sinusoidal signal of frequency 15 MHz.

\begin{figure}[htbp]
\centering
\includegraphics[width=10 cm]{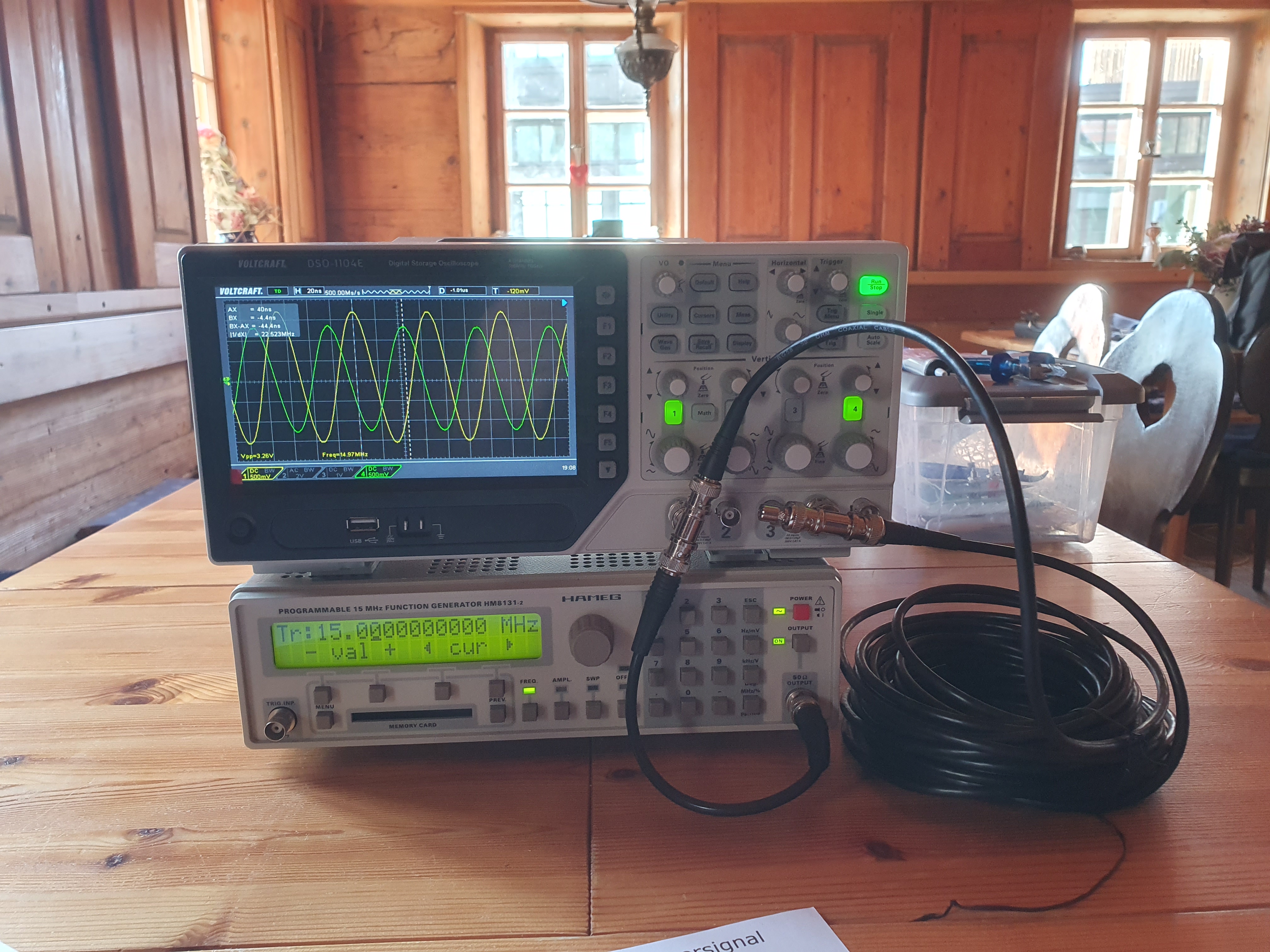}
\caption{Setup for measuring the signal speed along the coaxial cable.}
\label{fig:Lichtgeschwindigkeitsmessung}
\end{figure}

\fig{fig:Lichtgeschwindigkeitsmessung}) shows the input signal in yellow and the somewhat attenuated output signal, which is fed through the 10-meter cable, in green. The signal delay measured with the oscilloscope was $44\ns$. This corresponds to a signal speed in the cable of approximately $3/4$ of the speed of light in a vacuum:
\begin{equation}
c_\mathrm{cable} = \frac{L}{\Delta t} = \frac{10\m}{44\cdot 10^{-9}\s} = 2.27\cdot 10^8 \frac{\m}{\s} 
\end{equation}

The signal speed does not reach the speed of light because the electromagnetic waves propagate in the insulator between the cable shielding and the inner conductor. This results in a reduction by a factor of $\epsilon_\mathrm{r}^{-0.5}$ compared to the speed of light ($\epsilon_\mathrm{r}$: dielectric constant of the insulator). Further reduction may occur due to the attenuation of the signal when propagating along the cable.

\subsection{Measurements}
\subsubsection{Measurement with a load resistor of $150\Ohm$}

\begin{figure}[htbp]
\centering
\includegraphics[width=10 cm]{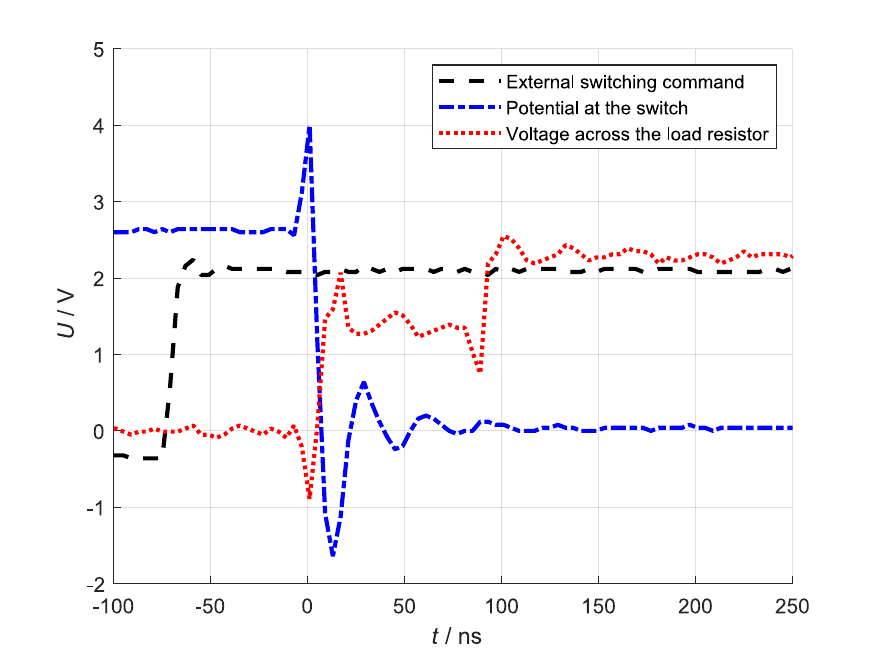}
\caption{Measurement results with a load resistor of $R=150\Ohm$: It can be seen that the voltage across the load resistor (red dotted line) changes at the same time as the switch is flipped (blue dash-dot line).
}
\label{fig:Messung-150Ohm}
\end{figure}

An initial measurement was carried out with a load resistance of $R=150\Ohm$. The measurement results are shown in \fig{fig:Messung-150Ohm}. 
After the signal generator triggers the switching process (black dashed line), it takes approximately $70\ns$ to $80\ns$ for the gate driver to respond. The result is a collapse of the potential at the switch, indicating the actual switching time (blue dash-dot line). 
The voltage across the load resistor (red dotted line) increases almost simultaneously with the collapse of the potential at the switch. It is therefore safe to conclude that at least something happens immediately after the switch is flipped. However, there is more to observe beyond this.

Two stages are clearly recognizable:
\begin{itemize}
\item In the first time period of $88\ns$ duration, the voltage across the load resistor increases almost instantly from $0$ to $1.5\V$ and remains at this value.
\item During the next $88\ns$, the voltage across the load resistor increases from $1.5\V$ to approximately $2.5\V$.
\item Further voltage steps, which should appear in theory, are not easily discernible.
\end{itemize}

\subsubsection{Measurement with a load resistor of $30\Ohm$}
\begin{figure}[htbp]
\centering
\includegraphics[width=10 cm]{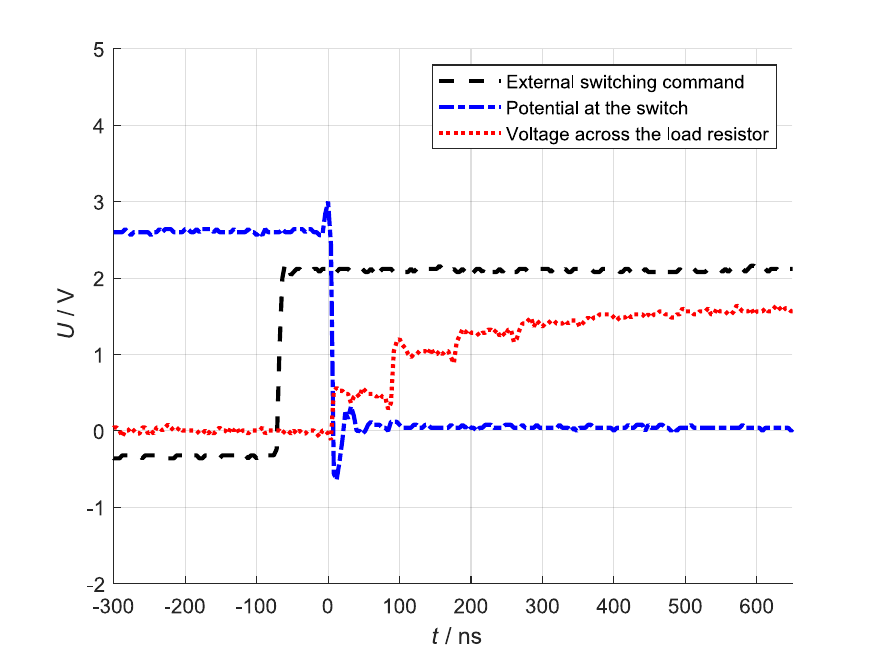}
\caption{Measurement results with a load resistor of $R=30\Ohm$: The final voltage builds up in steps, each with a duration of $88\ns$.
}
\label{fig:Messung-30Ohm}
\end{figure}

When the same experiment is repeated with a load resistor of $30 \Ohm$, it can be observed that the voltage increases in several steps with a duration of $88\ns$, each (see \fig{fig:Messung-30Ohm}). The dotted red line representing the voltage across the load resistor, shows at least $5$ distinguishable voltage steps. However, note that the measured voltages are lower than expected because the supply voltage collapses under load. The $47\uF$ electrolytic capacitor actually used to buffer the supply voltage is too slow, and the $100\nF$ capacitor wired in parallel is too small to maintain proper operation with such small load resistances (see \sect{sec:Voltage source} for more details).

\subsection{Explanation of the Measurement Results}

\subsubsection{Why does the voltage arrive at the resistor immediately?}
As can be seen from the measurement data (\fig{fig:Messung-150Ohm} and \fig{fig:Messung-30Ohm}), the voltage at the load resistor changes virtually without any delay after the switch is flipped.  

A question that many might ask about the outcome of the experiment is: \qq{How is it possible that at the same moment when a charge is fed into the inner conductor, the same amount of charge comes out of the shielding connection without any delay? Wouldn't this require a chain of electrons to be pushed one after another from the input of the coaxial cable to its end $10\m$ away and then back again, so that a period of $2 \times 44 \ns$ (corresponding to twice the signal transit time along the cable) elapses before anything happens at all at the load resistor?}

\begin{figure}[htbp]
\centering
\includegraphics[width=12 cm]{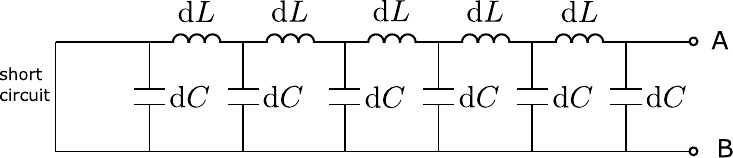}
\caption{Model of a two-wire transmission line.}
\label{fig:Wellenleitung}
\end{figure}

In fact, this is not the case. The reason is that the inner conductor and the outer conductor (shielding) of a coaxial cable -- as with any two-wire cable -- are capacitively coupled to each other. So if a charge is applied to the inner conductor, a correspondingly large countercharge is induced on the outer conductor and vice versa. As the inner and outer conductors are only a few millimeters apart in a coaxial cable, both processes occur almost instantaneously. 

\fig{fig:Wellenleitung} shows how a lossless coaxial cable, which is shorted on one side, is modeled in the theory of transmission lines. It consists of an infinite number of differentially small inductors and capacitors that are connected as shown. If you apply a voltage between the terminals A and B, the capacitors are sequentially charged from right to left. Due to the behavior of the inductors, there is a time delay between the charging of one capacitor and its successor. Ultimately, the theory of transmission lines describes the process occurring in the coaxial cable as voltage and current waves propagating along the cable.

The theory of transmission lines is used in the following to explain the behavior of the circuit in more detail. For an introduction to the topic, the reader may refer to the following Wikipedia entries: \href{https://en.wikipedia.org/wiki/Transmission_line}{Transmission Line} and \href{https://en.wikipedia.org/wiki/Telegrapher\%27s_equations}{Telegrapher's Equation}. To explore the topic further, the book by Steer \citep{Steer} can be read. It is freely available online. 

\subsubsection{Calculation of the voltage at the load resistor}
\begin{figure}[htbp]
\centering
\includegraphics[width=17 cm]{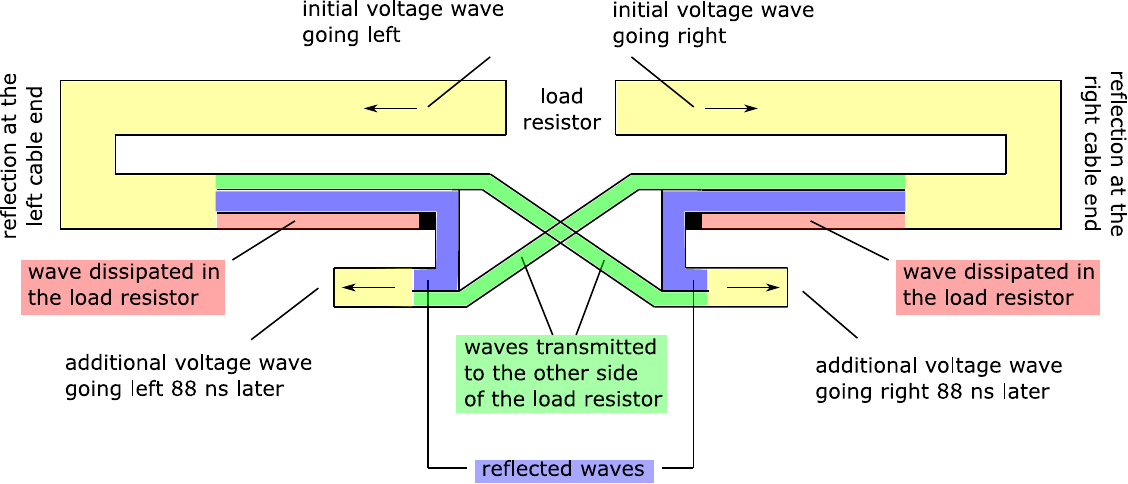}
\caption{The diagram shows which additional wave components are generated when the reflected waves hit the load resistor. Note that the waves can have either a positive or a negative sign. This is true in particular for the waves reflected at the cable ends, which have the same amplitude but the opposite sign compared to the incoming waves.}
\label{fig:Energy-flow}
\end{figure}

To understand the behavior of the circuit quantitatively, we proceed as follows (see \fig{fig:Energy-flow}): 
\begin{enumerate}
\item \textit{Calculation of the initial state:} First, we examine the initial state at time $t=0$. It can be calculated based on a circuit model with concentrated elements, where the coaxial cable is represented by a resistor with a resistance corresponding to the wave impedance of the cable.

Starting from this initial state, the amplitudes of the two voltage waves are calculated, which are emitted from the load resistor toward the ends of the coaxial cables. The potentials at both terminals of the resistor remain constant while the voltage waves propagate along the cables, reflect at the cable ends, and return to the load resistor after a propagation time of $T=88\ns$. $T$ represents twice the transit time of a signal along a $10\m$ cable (see \sect{sec:Lichtgeschwindigkeit}).

\item \textit{Wave propagation and superposition} Upon arrival, each of the waves returning to the load resistor is partially transmitted, partially reflected, and partially dissipated as heat inside the load resistor. We calculate the amplitudes of the transmitted and reflected waves and then determine the voltages at time $t=T^+$. The plus sign in $t=T^+$ indicates that the situation immediately after the arrival of the reflected waves is considered.
\item \textit{New initial state} The newly calculated state is then taken as the starting point for the next iteration, \hbox{i.\,e.} we use the updated potentials at the left and the right terminals of the load resistor at time $t=T^+$ as the initial state to calculate these potentials at time $t=2\cdot T^+$. 
\end{enumerate}
In this way, the time function of the electrical potentials at the left and right terminals of the load resistor can be calculated iteratively, step by step.

In the following, the calculation for a load resistance of $150\Ohm$ is carried out as an example. 

\begin{figure}[htbp]
\centering
\includegraphics[width=10 cm]{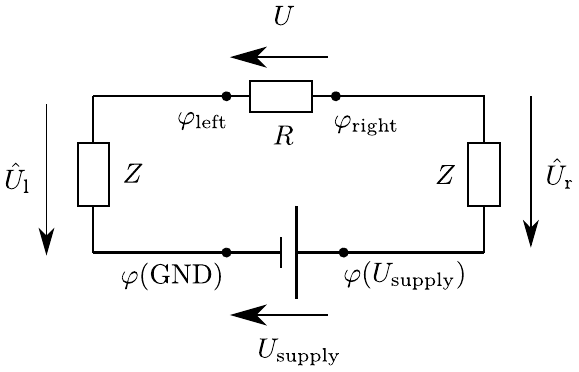}
\caption{Equivalent circuit for the first $88\ns$.}
\label{fig:Ersatzschaltbild}
\end{figure}

\paragraph{Initial State}
At the first moment of connection to the circuit, a coaxial cable (like any two-wire conductor) behaves like a resistor with the value of the so-called wave impedance $Z$ were connected. In our experiment, we used a coaxial cable with $Z=50\Ohm$ and a load resistance of $R=150\Ohm$. So, for the calculation of the initial state between $t=0$ and $t=T^{-}$, we consider the experimental setup to behave like the circuit shown in \fig{fig:Ersatzschaltbild}. 

Using the voltage divider rule, the values for the potentials at the terminals of the load resistor and the voltage across the resistor between $t=0^+$ and $t=T^{-}$ are as follows:
\begin{eqnarray}
\varphi_\mathrm{left} & = & \frac{Z}{2\cdot Z+R} \cdot U_\mathrm{supply} = \frac{50 \Ohm}{2\cdot 50 \Ohm + 150 \Ohm}\cdot 2.5\V = 0.5\V \\ 
\varphi_\mathrm{right} & = & \frac{Z+R}{2\cdot Z+R} \cdot U_\mathrm{supply} = \frac{50 \Ohm + 150\Ohm}{2\cdot 50 \Ohm + 150 \Ohm}\cdot 2.5\V = 2\V \\
U 									  & = & \varphi_\mathrm{right}-\varphi_\mathrm{left} = 2\V - 0.5\V = 1.5\V.
\end{eqnarray} 

The initial values of the voltage waves propagating to the left and to the right are thus
\begin{eqnarray}
\hat U_\mathrm{left}  & = & \varphi_\mathrm{left} - \varphi(\mathrm{GND}) = 0.5\V - 0\V = 0.5\V\\
\hat U_\mathrm{right} & = & \varphi_\mathrm{right} - \varphi(U_\mathrm{\mathrm{supply}}) = 2\V - 2.5\V = -0.5\V 
\end{eqnarray}

The results agree well with the measurements shown in \fig{fig:Messung-150Ohm}. For plausibility, note that the energy dissipated in the $50\Ohm$ resistor $Z$ in the circuit model is, in reality, the energy that propagates along the coaxial cable to the cable ends.

\paragraph{Step from $t=n\cdot T^{+}$ to $t=(n+1) \cdot T^{+}$} 
What happens, now, when the wave arrives at the cable ends and finally returns to the load resistor $R$? 

When a wave propagates from a transmission line with impedance $Z_1$ to a transmission line (or circuit) with impedance $Z_2$, it is reflected with the reflection factor 
\begin{equation}
\underline r=\frac{Z_2-Z_1}{Z_2+Z_1}
\end{equation}
and transmitted with the transmission factor 
\begin{equation}
\underline t=1+\underline r=\frac{2 \cdot Z_2}{Z_2+Z_1}
\end{equation}

When applied to the voltage wave in the coaxial cable running toward the shorted cable ends ($Z_1=Z$ and $Z_2=0$), the reflection factor equals 
$\underline r = (0-Z)/(0+Z) = -1$, and the transmission factor is $\underline t=1+\underline r=0$. This means that the waves that are reflected at the cable ends return to the resistor with the opposite sign, but otherwise unchanged.

\begin{figure}[htbp]
\centering
\includegraphics[width=7 cm]{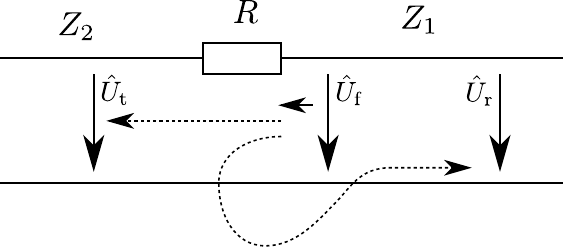}
\caption{A voltage wave propagating from right to left is transmitted and reflected at an inhomogeneity. The indices represent the forward wave (f), transmitted wave (t), and reflected wave (r), respectively.}
\label{fig:Reflection}
\end{figure}

\fig{fig:Reflection} demonstrates the situation when a wave returns to the load resistor $R$ after being reflected at the right-hand cable end. In this case, the amplitudes of the transmitted and reflected waves are:
\begin{eqnarray}
\hat U_\mathrm{r}&=&\underline r\cdot \hat U_\mathrm{f} = \frac{(R+Z_2)-Z_1}{(R+Z_2)+Z_1} \cdot \hat U_\mathrm{f} \underbrace{=}_{Z_1=Z_2} \frac{R}{R+2\cdot Z} \cdot \hat U_\mathrm{f}, \textrm{~~~~and} \\
\hat U_\mathrm{t}&=&(1+\underline r)\cdot \hat U_\mathrm{f} = \frac{2\cdot Z_2}{Z_2+(R+Z_1)} \cdot \hat U_\mathrm{f} \underbrace{=}_{Z_1=Z_2} \frac{2\cdot Z}{2\cdot Z+R} \cdot \hat U_\mathrm{f}.
\end{eqnarray} 

When given the variables $\varphi_\mathrm{left}(n)$, $\varphi_\mathrm{right}(n)$, $\hat U_\mathrm{left}(n)$ and $\hat U_\mathrm{right}(n)$ at the time $t=n\cdot T^{+}$, the corresponding variables at time $t=(n+1)\cdot T^{+}$ are calculated by:
\begin{eqnarray}
\hat U_\mathrm{left}(n+1) &=& \hat U_\mathrm{left}(n) + (-1) \cdot \hat U_\mathrm{left}(n) + (-1) \cdot \hat U_\mathrm{left}(n) \cdot \underline r + (-1) \cdot \hat U_\mathrm{right}(n) \cdot \underline t \label{eq:left}\\
\hat U_\mathrm{right}(n+1) &=& \hat U_\mathrm{right}(n) + (-1) \cdot \hat U_\mathrm{right}(n) + (-1) \cdot \hat U_\mathrm{right}(n) \cdot \underline r + (-1) \cdot \hat U_\mathrm{left}(n) \cdot \underline t \label{eq:right}
\end{eqnarray}

Each of the terms for the left amplitude (\eq{eq:left}) has the following meaning: 
\begin{itemize}
\item $\hat U_\mathrm{left}(n)$: the initial voltage at the current position, which is maintained because the voltage supply is still switched on,
\item $(-1) \cdot \hat U_\mathrm{left}(n)$: the wave arriving back from the left end of the wire, which is considered with a minus sign due to the reflection at the shorted cable end,
\item $(-1) \cdot \hat U_\mathrm{left}(n) \cdot \underline r$: the reflection of the wave coming from the left  end of the cable at the resistor $R$, where the wave is now running back to the cable ends,
\item $(-1) \cdot \hat U_\mathrm{right}(n) \cdot \underline t$: the transmission of the wave coming from the other side of the resistor $R$.
\end{itemize}

The terms for \eq{eq:right} are likewise (see \fig{fig:Energy-flow} for a visualization of the energy flow).

Adding the ground and battery potentials yields the electrical potentials on the left and the right side, respectively:
\begin{eqnarray}
\varphi_\mathrm{left}(n+1) &=& \varphi(\mathrm{GND}) + \hat U_\mathrm{left}(n+1) = 0\V +  \hat U_\mathrm{left}(n+1),  \\
\varphi_\mathrm{right}(n+1) &=& \varphi(U_\mathrm{supply}) + \hat U_\mathrm{right}(n+1) = 2.5\V + \hat U_\mathrm{right}(n+1) \textrm{~~~ and} \\
U(n+1) &=& \varphi_\mathrm{right}(n+1) - \varphi_\mathrm{left}(n+1) = 2.5\V + \hat U_\mathrm{right}(n+1) - \hat U_\mathrm{left}(n+1)
\end{eqnarray}

\begin{figure}[htbp]
\centering
\includegraphics[width=12 cm]{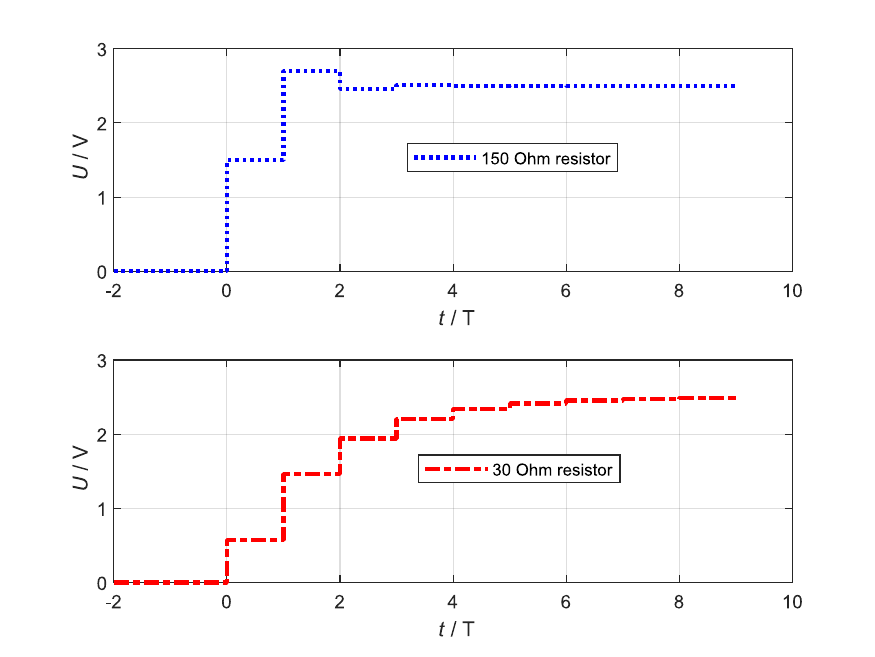}
\caption{Calculated voltage across the load resistor of $R=150\Ohm$ (blue dotted line) and $R=30\Ohm$ (red dash-dot line).}
\label{fig:Calculated-voltage-150}
\end{figure}

Using the variables associated with the measurement, the time development of the voltage across the load resistor was calculated in the \textsc{MATLAB}-script shown in \app{app:Voltage}. The results are depicted in \fig{fig:Calculated-voltage-150}.

\subsection{Simulation of the Circuit Behavior}
\begin{figure}[htbp]
\centering
\includegraphics[width=12 cm]{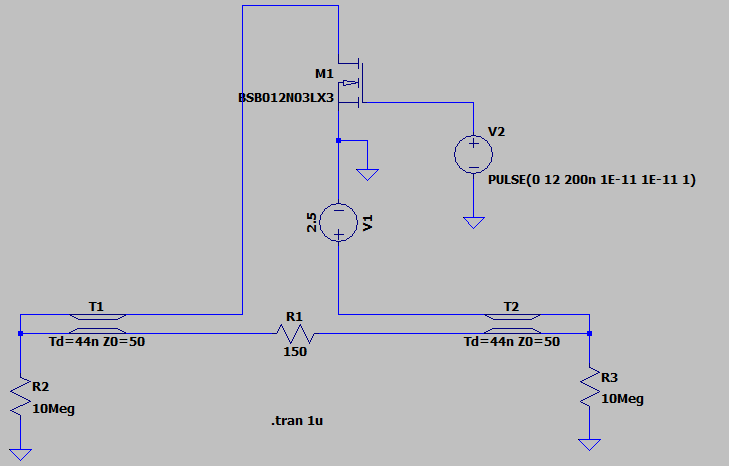}
\caption{Circuit with a load resistor of $150\Ohm$ as an LTSpice simulation network.}
\label{fig:Simulation-Circuit}
\end{figure}

Using the LTSpice circuit simulator \citep{LTSpice}, the circuit can be simulated accordingly by using the "tline" component (transmission line) as a model for the coaxial cable. A suitable setup is shown in \fig{fig:Simulation-Circuit}. The $1\MOhm$ resistors $R_2$ and $R_3$ ensure the numerical stability of the solver. They do not serve a functional purpose.  

It turns out that, with the exception of some transients at the switching times, the LTSpice simulations yield virtually identical results to the calculations shown in \fig{fig:Calculated-voltage-150}. 

If different load resistors or a cable with a different wave impedance are used, the basic behavior of the circuit does not change. However, depending on the properties of the components, it is possible that a significantly longer or shorter time will pass before the steady-state conditions are reached. 

\section{Conclusion}
\label{sec:Conclusion}
The claim made in Veritasium's video that the energy essentially travels from the voltage source to the load resistor without a detour through the long cables is highly counterintuitive for many students. Many believe that the electrical energy is attached to the charges and moves with them from the voltage source to the load.

The aim of this paper is therefore to provide colleagues in schools and universities with the opportunity to experimentally correct this widespread misconception using a low-cost measurement setup.

The fact that the electrical power arrives at the load resistor immediately after flipping the switch is clearly visible in the measurement. Its undeniable consequence is that the energy must have taken the direct route through the air rather than traveling back and forth along the cables.

This experimental refutation is all the more important, as the concept of the Poynting vector can only be discussed with very advanced students.

It is to the credit of the Veritasium channel that it has packaged these profound questions in an extremely appealing and exciting video, thus inspiring many learners to engage intensively with physics.

\newpage
\appendix   

\section{PCB Description and Parts}
\label{sec:Sources}

\begin{figure}[h]
\centering
\includegraphics[width=12cm]{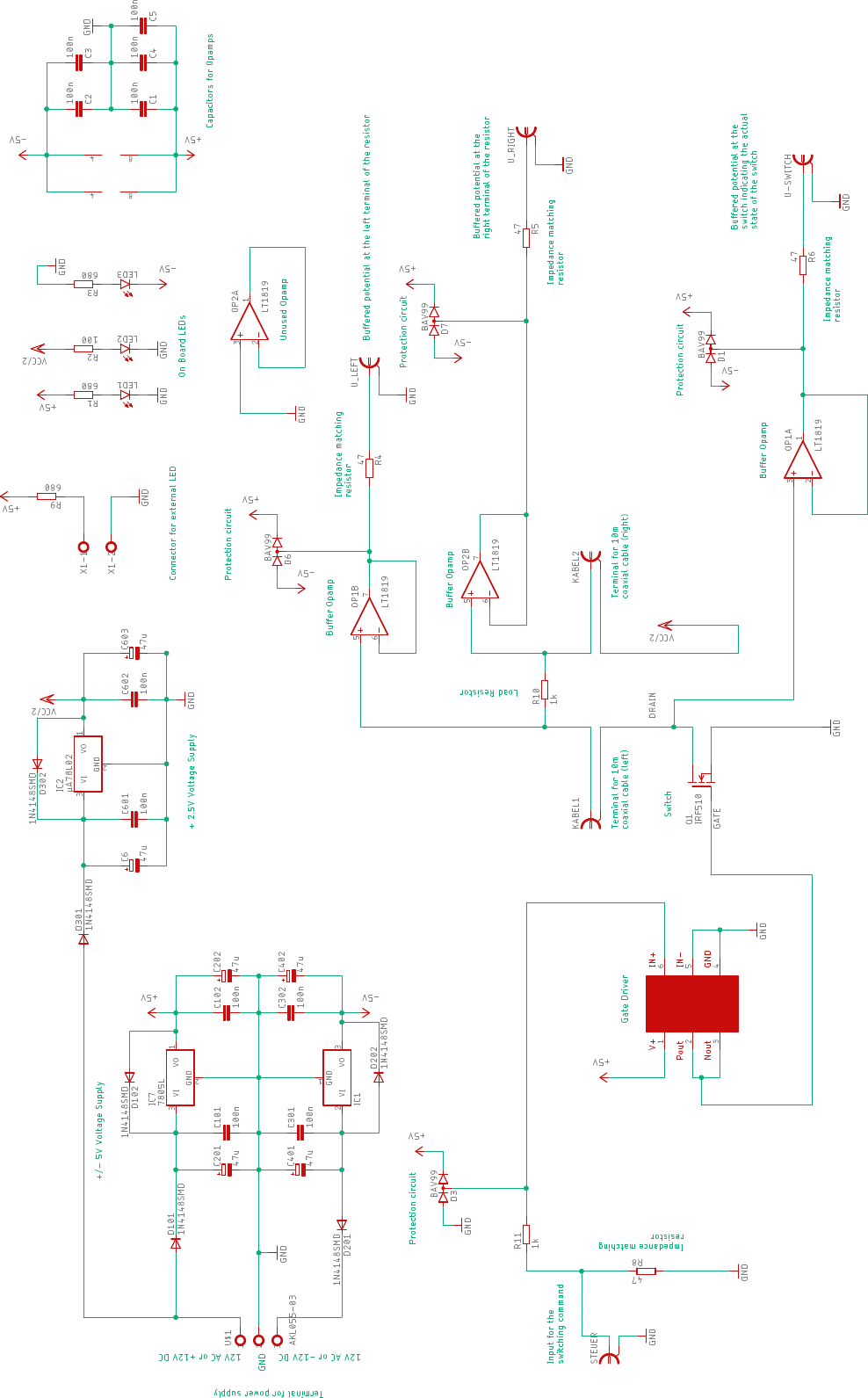}
\caption{Overall circuit}
\label{fig:Gesamtschaltplan}
\end{figure}

The overall circuit of the PCB is found in \fig{fig:Gesamtschaltplan}. 
The Eagle PCB layout files\citep{Eagle} and Gerber production files can be sent by email upon request.


All relevant components and the order numbers from companies (\href{https://www.reichelt.de/}{Reichelt}) and Mouser (\href{https://www.mouser.de/}{Mouser}) are listed below. The components listed first have been tested by us; the components listed second and third (introduced by the word \qq{or}) have not been tested but should be usable without any issues.

\begin{table}[htbp]
\caption{Components provided by Reichelt}
\begin{tabularx}{15cm }{X |X}
\toprule[2pt]
Aluminum housing (optional)                                     & Series resistor $5\V$ LED\\
1 x GEH EG-2                                                    & 2 x SMD-0805 680 \\
\midrule[1pt]
Connection terminal for LED                                     & Series resistor $2.5\V$ LED \\
1 x AKL 101-02      ~~~or~~~  1 x AKL 055-02                    & 1 x SMD-0805 100\\ 	
\midrule[1pt]
Connectors for voltage supply                                   & Linear regulators \\
1 x BIL 20 BL, 1 x BIL 20 RT, ~and~                      
& 1 x µA 78L05, 1 x µA 79L05, and\\
1 x BIL 20 SW                                                   & 1 x µA 78L02\\
\midrule[1pt]
Connectors on PCB (3 poles)                                     & Double diodes\\
1 x AKL 101-03    ~~~or~~~  1 x AKL 055-03                      & 4 x BAV 99\\ 	
\midrule[1pt]
BNC connectors, angled, $50\Ohm$                                & Load resistors\\
6 x UG 1094W1                                                   & e.g. 3 x METALL 330\\
\midrule[1pt]
BNC cables, $10\m$ long, $50\Ohm$                               & Resistors for impedance matching\\
2 x KBK 110 10M                                                 & 4 x SMD-0805 47,0\\ 
\midrule[1pt]
BNC resistors/adapters                                          & Other resistors\\ 
UG 88-50 ~~~ UG 914-U ~~~ UG 274-U                              & 1 x SMD-0805 1,00K\\
\midrule[1pt]
LEDs for status indication (red)                                & Electrolytic capacitors $47\uF$, grid $2.5\mm$ \\
3 x SMD-LED 0805 RT                                             & 6 x RAD FR 47/25 or 6 x RAD FR 47/63 \\
\midrule[1pt]
Diode 1N4148 SMD                                                & $100\nF$ capacitors, SMD\\
6 x 1N 4148 SMD                                                 & 11 x X7R-G0805 100N      or \\
                                                               & 11x KEM X7R0805 100N  \\
\bottomrule[2pt]
\end{tabularx}
\label{tab: Reichelt}
\end{table}

\newpage

\newpage

\begin{table}[htbp]
\caption{Components provided by Mouser}
\begin{tabularx}{15cm }{X | X}
\toprule[2pt]
Operational amplifiers                                         & Gate driver\\
2 x 584-LT1819CS8\#PBF  or                                      & 1 x 700-MAX5048CAUT+T or \\
2 x 584-LT1819IS8\#PBF                                          & 1 x 595-LM5114AMF/NOPB \\
\midrule[1pt]
NFET (switch)                                                   & BNC short \\
1 x 942-IRFZ24NPBF  or                                          & 2 x 523-202114\\ 	
1 x 844-IRF510PBF  or                                           & ~\\
1 x IRF 510 (Reichelt)                                          & ~\\
\bottomrule[2pt]
\end{tabularx}
\label{tab: Mouser}
\end{table}

\section{Design Considerations}
\label{sec:Design-considerations}
The following discusses various aspects of the circuit and design considerations for a better understanding.

\subsection{Voltage source}
\label{sec:Voltage source}

\begin{figure}[htbp]
\centering
\includegraphics[width=10 cm]{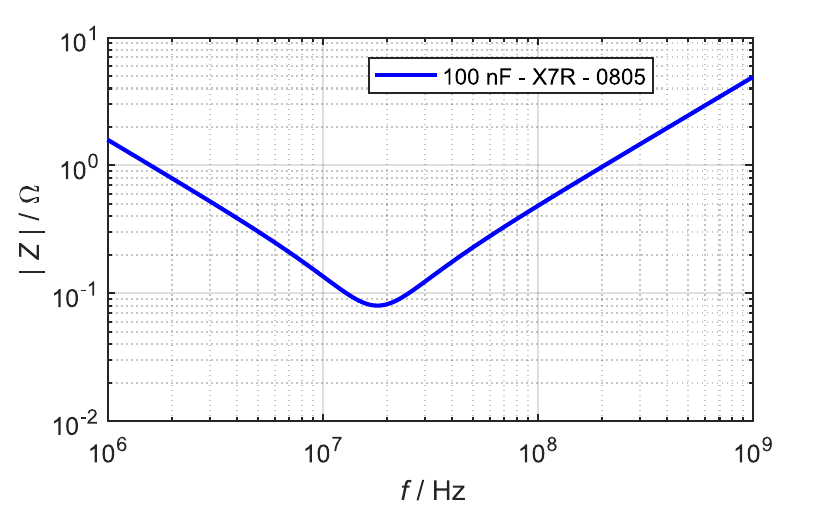}
\caption{Typical resonance curve of a $100\,\mathrm{nF}$ capacitor made of the material X7R in SMD technique. \citep{AVX}}
\label{fig:Resonanz}
\end{figure}

To guarantee a constant voltage, a linear voltage regulator with a supply voltage of $2.5\V$ (78L02 by Texas Instruments) was used and supplied with two capacitors each at its input and output: a $47\uF$ electrolytic capacitor and a $100\nF$ capacitor in an 0805 SMD package. The SMD capacitor at the output is an important component of the supply circuit for this experiment. It is much faster than the electrolytic capacitor and therefore supplies the charge in the critical time interval immediately after the switch is flipped. As you can see in \fig{fig:Resonanz}, capacitors in high-frequency circuits are not just considered to be capacitors, but rather an RCL-series resonance circuit, where $R$ is the resistance and $L$ is the inductance of the conductors on the board and the inner leads of the component. With a resonance frequency of about $20 \MHz$, this capacitor remains suitable for the present application, although a higher resonance frequency would be preferable. In a new design, I would consider using a parallel connection of several capacitors with the NP0 dielectric instead of the X7R dielectric used here. In practice, it is important that the capacitor is installed in the immediate vicinity of the FET in order to provide the required charge for the circuit quickly enough. 

\subsection{Stripline and BNC cable}
\label{sec:Stripline-and-Bnc}
As a model for the long cables of the Veritasium experiment, two $10\m$ long BNC cables with a wave impedance of $50\Ohm$ are used and have been short-circuited at their ends.
In order to minimize signal reflections at the transition between the PCB and the BNC cable, a stripline with the same wave impedance as the BNC cable was implemented on the PCB. For standard PCBs made of FR4 material ($\mu_r = 4.3$; dielectric thickness: $1.55\mm$) and a conductor track thickness of $35\um$, this requires a conductor track width of $3\mm$. The return conductor was designed as a surface with extensions much larger than the stripline. 

\subsection{FET and FET driver}
An N-channel field effect transistor (IRFZ24N) in a TO-220 package serves as a switch. This transistor can be switched with a logic-compatible gate voltage of $2 \dots 4\V$. It has a drain-source resistance of only $70 \mOhm$ and can therefore be regarded as a good approximation of a short circuit. 
When the FET is driven with normal laboratory equipment such as a signal generator or a microcontroller board (Arduino), the switching process of this FET takes much longer than the signal propagation time along a $10\m$ cable. This is due to its high gate capacitance of $13.3 \nC$, which has to be loaded with charge before the FET actually switches.
For this reason, a so-called gate driver (model: MAX5048 from Analog Devices) was connected between the generator of the switching pulse and the gate of the FET. This shortens the switching process in the FET to a duration of approximately $10\ns$, which is acceptably low compared with the measured propagation time of $88 \ns$ back and forth along the $10\m$ cable.  

\subsection{Decoupling circuit}
The voltage can be tapped at three crucial points and visualized with an oscilloscope:
\begin{itemize}
\item At the drain connection of the field-effect transistor,
\item at the left terminal of the load resistor, and
\item at the right terminal of the load resistor.
\end{itemize}
The potential at the drain connection of the field-effect transistor is used to detect the actual switching time of the transistor. Using a commercially available multi-channel digital oscilloscope, the potentials of the left and right terminals of the load resistor were measured and subtracted from each other in \textsc{MATH} mode, so that the voltage across the load resistor could be displayed on the oscilloscope. 
The comparison between the voltage across the load resistor and the potential at the drain connection allows a statement to be made about the delay between the switching operation at switch S and the increase in voltage at the load resistor (this corresponds to the “lighting of the light bulb” in the original experiment).  
The voltages at the three locations are tapped via a short conductor path, whereby the voltages are each connected to the high-impedance non-inverting input of an operational amplifier and amplified with the gain factor $v = 1$ (unity gain). This amplification is used for decoupling between the coaxial cable going “to the Moon” and the measurement cables to the oscilloscope. 
Without the amplifiers, the two measurement cables to the oscilloscope and any reflections at their cable ends would have to be taken into account, in addition to the two coaxial cables leading “to the Moon”, which would significantly complicate the evaluation and interpretation of the measurement results.

\begin{figure}[htbp]
\centering
\includegraphics[width=10 cm]{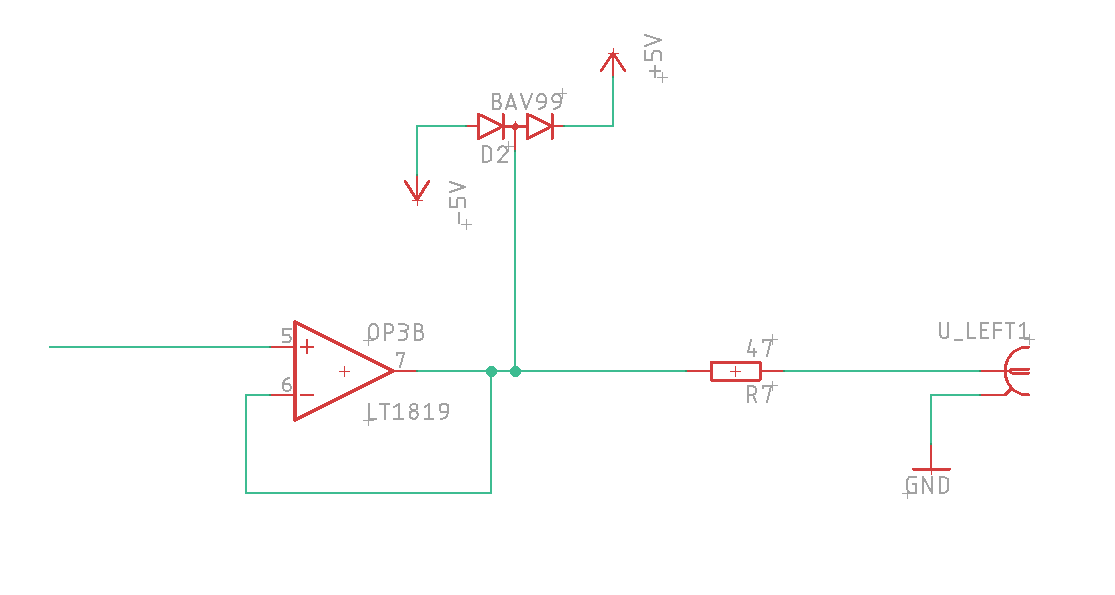}
\caption{Amplifier circuit for the switching voltage and the potentials at the left and right terminals of the load resistor.}
\label{fig:Amplifier}
\end{figure}

The amplifier circuit is shown in \fig{fig:Amplifier}. The operational amplifier LT1819 by Analog Devices was used and wired as a unity gain amplifier. The decisive selection criteria for the specific model of the operational amplifier were the high gain-bandwidth product of approximately $400 \MHz$, the high slew rate with possible voltage rise rates at the output of $2500 \V /\us$ and the stability with a gain factor of $v = 1$. An operational amplifier that is not unity-gain-stable would work as an oscillator with the realized circuitry and continuously output high-frequency oscillations with maximum amplitude.
For impedance matching to the coaxial cable leading to the oscilloscope, the output is connected in series with a $47\Ohm$ resistor. This resistor value corresponds approximately to the $50 \Ohm$ wave impedance of the coaxial cable and prevents wave reflections at the transmitter. The double diode BAV99 protects the output of the operational amplifier from any electrostatic charges that could be introduced via the BNC connection.

\newpage
\section{Matlab Script for Voltage at Load Resistor}
\label{app:Voltage}
\begin{table}[h!]
\begin{lstlisting}[caption={Matlab algorithm for calculating the voltage at the load resistor} \label{lst: Voltage}]
N=10;               % number of time intervals to plot   
Z=50;               % wave impedance
R=150;              % load resistance
Usupply = 2.5;      % supply voltage

Ul=Z/(2*Z+R)*Usupply;                   % initial ampl. voltage left
Ur=(Z+R)/(2*Z+R)*Usupply-Usupply        % initial ampl. voltage right 
r = R/(R+2*Z);                          % reflection factor 
t = 2*Z/(2*Z+R);                        % transmission factor 

for i=1:N-1,
  Ul(i+1)=Ul(i)-Ul(i)-r*Ul(i)-t*Ur(i);  % new amplitude left 
  Ur(i+1)=Ur(i)-Ur(i)-r*Ur(i)-t*Ul(i);  % new amplitude right
end

t = -2:N-1;                             % time variable
signal = 2.5+Ur-Ul;                     % voltage signal
signal=[0,0,signal];                    % signal for t<0

figure(1);
subplot(2,1,1);
stairs(t,signal,'b:','linewidth',2);    % plot voltage curve
xlabel('{\it t} / T');                          
ylabel('{\it U} / V');   
legend('150 Ohm resistor');
grid on;
\end{lstlisting}
\end{table}

\begin{acknowledgments}
Special thanks go to Mr. Uwe Kopte from Geschwister-Scholl Gymnasium in Löbau, who has been promoting gifted students in the field of physics at the specialist camp in Seifhennersdorf for several decades and has created an inspiring environment for both students and teachers. Without his dedication, these experiments would have never been realized. Many thanks also to Johann Höpfner and Richard Schlossarek for their valuable advice during the proofreading process.
\end{acknowledgments}

\section*{Author Declarations section}
The author has no conflicts to disclose.

\section*{Credit Line}
The following article has been submitted to the \qq{American Journal of Physics}. After it is published, it will be found at \href{https://publishing.aip.org/resources/librarians/products/journals/}{Link}.

Copyright (2024) Michael Lenz. This article is distributed under a Creative Commons Attribution-NonCommercial-NoDerivs 4.0 International (CC BY-NC-ND) License. \href{https://creativecommons.org/licenses/by-nc-nd/4.0/}{https://creativecommons.org/licenses/by-nc-nd/4.0/}

\end{document}